\documentclass[aps,prd,12point,twocolumn,nofootinbib,showpacs,superscriptaddress]{revtex4-1}
\def\theequation{\arabic{section}.\arabic{equation}}
\usepackage{psfrag}
\usepackage{subfigure}
\usepackage{color}
\usepackage{mathrsfs}
\usepackage{graphicx}
\usepackage{amssymb, bm}
\usepackage{amsmath, amsthm}
\usepackage{epstopdf}
\usepackage{hyperref}
\usepackage{enumerate}
\usepackage{longtable}

\newcommand{\be}{\begin{equation}}
\newcommand{\ee}{\end{equation}}

\begin{document}
\def\theequation{\arabic{section}.\arabic{equation}}


\title{Analogy between freezing lakes and the cosmic radiation era}


\author{Valerio Faraoni}
\email[]{vfaraoni@ubishops.ca}
\affiliation{Department of Physics \& Astronomy, Bishop's University\\
2600 College Street, Sherbrooke, Qu\'ebec, Canada J1M~1Z7
}


\begin{abstract}

An equation describing a one-dimensional model for the freezing of lakes 
is shown to be formally analogous to the Friedmann equation of cosmology. 
The analogy is developed and used to speculate on the change between two 
hypothetical ``spacetime phases'' in the early universe.

\end{abstract}


\maketitle

\section{Introduction}
\label{sec:1}
\setcounter{equation}{0}

The study of the freezing of water bodies in the natural 
environment has a 
long history \cite{LF1}-\cite{LF10}, appearing also in pedagogical 
\cite{LFpedagogical1, LFpedagogical2, Vollmer} and popular 
\cite{LFpopular} literature. The realistic problem of freezing of lakes in 
winter or in cold (for example, mountainous or polar) regions is difficult 
when factors such as the variability of atmospheric conditions, 
boundaries, and chemical impurities are taken into account 
\cite{LF1}-\cite{LF10}, but it can be simplified considerably and reduced 
to a one-dimensional model under certain assumptions, which are best 
spelled out in the pedagogical literature \cite{Vollmer}. These 
assumptions are: the lake covers a large area and effects 
at the margins can be neglected; the lake is isolated, {\em i.e.}, not 
connected to other 
lakes, rivers, or bodies of water; there are no chemical impurities (the 
lake contains only freshwater); the lake is deep (in practice, depths 
larger than $\sim 10$~m and short periods of cold weather are considered, 
but this assumption can be relaxed to allow for lakes that freeze to the 
bottom or are permanently frozen, or for long periods of colder weather 
as in 
polar or  subpolar climates or in regions at high elevation). One further 
neglects 
the geothermal input from the lake bottom, and solar radiation during the 
day.

The variables and parameters of the model then include \cite{Vollmer} the 
ice density $\rho_{ice}$, the thermal conductivity of ice $\lambda_{ice}$, 
the latent heat of fusion of water $L_f$, the ice thickness (measured from 
the surface) $z$, and the water and air temperatures $T_1$ and $T_3$. The 
heat losses from the lake ice to the atmosphere due to convection and 
radiation are simplified and described by a single heat flux density 
linear in the difference between air and ice temperatures \cite{Vollmer} 
and described by a single heat coefficient $h$. This is the main 
simplification of the model that makes the phenomenon tractable 
analytically, since both of these fluxes are in reality non-linear and 
convection is notoriously difficult to model due to changing air 
conditions, wind, {\em etc.} We will nevertheless adopt the simplified 
model of Ref.~\cite{Vollmer}. In this model, equating the heat flux 
density from the ice surface to the atmosphere due to radiation and 
convection with the flux density from water to air due to conduction 
through the ice, a 
simple equation describing the time dependence of the ice thickness $z(t)$ 
is obtained \cite{Vollmer}:
\be
\frac{dz}{dt} = \frac{ 2 \left( T_1-T_3\right)}{\rho_{ice} L_f} \, 
\frac{1}{ \frac{1}{h}+\frac{z}{\lambda_{ice}} } \,.\label{Vollmer}
\ee 
For our purposes, it is convenient to rewrite this equation in a  
different form. By introducing 
\begin{eqnarray}
\alpha &\equiv & \frac{2(T_1-T_3)}{\rho_{ice}L_f \lambda_{ice}} \,, \\
&&\nonumber\\
y &\equiv & \frac{z}{\lambda_{ice}} \,, \;\;\;\;\;\
y_0 \equiv \frac{1}{h}  \,,\\
&&\nonumber\\
s &\equiv & y+y_0 = \frac{z}{\lambda_{ice}} +y_0  \,, \\
\end{eqnarray}
and squaring it, Eq.~(\ref{Vollmer}) is rewritten as 
\be
\left( \frac{\dot{s}}{s} \right)^2 =\frac{\alpha^2}{s^4} \,. 
\label{Vollmerfors}
\ee
This equation is analogous to the Friedmann equation of relativistic 
cosmology for a spatially flat Friedmann-Lema\^itre-Robertson-Walker 
(FLRW) universe filled with blackbody radiation, as discussed in the next 
section. Indeed, the Friedmann equation, which resembles an energy 
conservation equation for a conservative mechanical system, lends itself 
to analogy with equations arising in the study of many different and 
completely unrelated physical systems, ranging from particles in 
one-dimensional motion \cite{particle1}-\cite{particle5} to optical 
systems 
\cite{Chen15a, Chen15b}, condensed matter systems \cite{condensed1, 
condensed2, condensed3, condensed4}, the transverse profiles of glacial 
valleys \cite{Chen15a, 
Chen15b, facets}, and equilibrium 
beach profiles \cite{beach}.

\section{Cosmological analogy}
\label{sec:2}
\setcounter{equation}{0}

Let us recall the essential equations of FLRW cosmology in order to 
develop the analogy with the freezing of lakes. We follow the notation of 
Refs.~\cite{Wald, Carroll}, in which the speed of light is unity and $G$ 
is Newton's constant.

\subsection{FLRW cosmology: basics}

The geometry of a  spatially homogeneous and  isotropic 
universe is necessarily given by the four-dimensional FLRW line 
element which, in comoving polar coordinates $\left(t, r, \theta, \varphi 
\right)$, reads
\be
ds^2 = -dt^2 +a^2(t) \left[ \frac{dr^2}{1-Kr^2} +r^2 \left( d\theta^2 + 
\sin^2 \theta \, d\varphi^2 \right)\right] \,. \label{eq:10}
\ee
In this class of solutions of the Einstein field equations of general 
relativity (GR), the 
scale factor $a(t)$ embodies the expansion history of the universe. 
According to the sign of the constant $K$, which describes the constant 
curvature of the 3-dimensional spatial geometries obtained by setting  
$dt=0$, the FLRW line 
element~(\ref{eq:10}) describes closed universes (for $K>0$), or Euclidean 
spatial 
sections (when $K=0$), or hyperbolic 3-spaces (if $K<0$), respectively 
\cite{Wald, Carroll, Liddle, KT}. This classification includes all the 
possible FLRW geometries.

The matter content of the universe, which generates the spacetime 
curvature, is usually described by a perfect fluid with energy 
density $\rho(t)$ and isotropic pressure $P(t)$ related by an equation 
of state $P=P(\rho)$. The Einstein-Friedmann equations
satisfied by $a(t), \rho(t)$, and $P(t)$ are \cite{Wald, Carroll, Liddle, 
KT}  
\begin{eqnarray}
&&H^2 \equiv \left( \frac{\dot{a}}{a}\right)^2 =\frac{8\pi G}{3} \, \rho 
-\frac{K}{a^2} \,, \label{eq:11}\\
&&\nonumber\\
&&\frac{\ddot{a}}{a}= -\, \frac{4\pi G}{3} \left( \rho +3P \right) \,, 
\label{eq:12} \\
&&\nonumber\\
&& \dot{\rho}+3H\left(P+\rho \right)=0 \,,\label{eq:13}
\end{eqnarray}
where an overdot denotes differentiation with respect to the comoving 
time $t$ and 
$H(t)\equiv 
\dot{a}/a$ is the Hubble function \cite{Wald, Carroll, Liddle, KT}. Only 
two  of 
these three equations are independent: given any two, the third one can be 
derived from them. Without loss of 
generality, we adopt the Friedmann equation~(\ref{eq:11}) and the energy 
conservation equation~(\ref{eq:13}) as primary, while the acceleration 
equation~(\ref{eq:12}) is derived from them. 

\subsection{The analogy}

Equation~(\ref{eq:11}) with $K=0$ is formally the same as 
Eq.~(\ref{Vollmerfors}) ruling the thickness of ice in freezing lakes if 
we exchange the variables $\left( t, s(t) \right) \longrightarrow \left( 
t, a(t) \right)$. The analogy holds only if the energy conservation 
equation is also satisfied; this happens if a suitable cosmological fluid 
fills the analog universe. By comparing Eqs.~(\ref{Vollmerfors}) 
and~(\ref{eq:11}), we see that it must be
\be
\rho(t)=\frac{\rho_0}{ a^4(t)  } 
\,,\label{eq:14} 
\ee 
where $\rho_0 $ is a positive integration constant determined by the 
initial conditions. This relation is familiar in cosmology \cite{Wald, 
Carroll, Liddle, KT} and in blackbody thermodynamics ({\em e.g.}, 
\cite{Carter}). More in general, if the cosmic fluid 
satisfies the barotropic equation of state $
P=w\rho $ 
for a suitable constant $w$ (``equation of state 
parameter''), the conservation equation~(\ref{eq:13}) integrates to 
\be
\rho(a) = \frac{ \rho_0}{ a^{3(w+1)} }  \label{eq:16} \,.
\ee
Equations~(\ref{eq:14}) and~(\ref{eq:16}) then imply that the analogy 
between 
the freezing of lakes and cosmology holds if the analog universe is filled 
with blackbody radiation with equation of state parameter $ w =1/3 $ and 
energy density scaling as $\rho\sim a^{-4}$. The fact that the constant 
$\rho_0$ in Eq.~(\ref{eq:16}) is 
\be
\rho_0=\frac{3\alpha^2}{8\pi G} \label{rho0}
\ee
(as follows from Eqs.~(\ref{eq:14}), (\ref{eq:11}), and 
(\ref{Vollmerfors})) is fortunate 
because it implies that the analog fluid has always positive energy 
density and 
satisfies the energy conditions. This fact cannot be taken for granted in 
such a far-fetched analogy between physical phenomena that are so distant 
from each other.
 
The acceleration equation~(\ref{eq:12}) implies that the expansion of the 
radiation-dominated analog universe is always decelerated, $\ddot{a}<0$.

The analogy holds only if $K=0$ and 
\be
s(t)=\frac{z(t)}{\lambda_{ice}} +\frac{1}{h} 
\ee
is analogous to the scale factor $a(t)$, while $\dot{s}/s$ is analogous to 
the Hubble function $\dot{a}/a$. It is straightforward to verify  
that the energy conservation equation and the acceleration equation are   
satisfied with $w=1/3$ and $\rho(t)=\rho_0/a^4$ if $\rho_0$ is as in 
Eq.~(\ref{rho0}).

The solution of Eq.~(\ref{Vollmer}) with the initial condition $z(0)=0$ is 
\cite{Vollmer}
\be
z(t) = \lambda_{ice}\left[ \sqrt{ \frac{2\alpha t}{\lambda_{ice}} 
+\frac{1}{h^2} } -\frac{1}{h} \right] \,,
\ee
or
\be
s(t) =  \sqrt{ \frac{2\alpha t}{\lambda_{ice}} 
+\frac{1}{h^2} }  
\ee
with initial condition
\be
s(0)=\frac{1}{h} \equiv y_0 
\ee
(the well known square-root solution describes the radiation era of 
spatially flat FLRW universes filled with 
radiation \cite{Wald, Carroll, Liddle, KT}). 
Freezing begins at $t=0$ and $z=0$, corresponding to the 
finite value $a_0=1/h$ of the analog scale factor, while 
the Big Bang $a=0$ (or $s=0$) corresponds to 
\be
t_0= -\frac{\lambda_{ice}}{2\alpha h^2} \,, \;\;\;\;\;\;\;
z_0=-\frac{\lambda_{ice}}{h} \,, \;\;\;\;\;\;\;\; y=-y_0 \,.
\ee 

It is interesting to speculate what the analogy with freezing lakes could  
imply if the state of the universe corresponds to a ``phase of gravity'' 
and the liquid-solid phase transition of water has some analogue in 
gravity at the high energies and temperatures found in the early universe. 
In string gas cosmology, the early universe is regarded as a gas of 
strings and there is a phase transition at the Hagedorn temperature. 
Correction terms to the low-energy effective 
action of  string theory dominate
in the Hagedorn phase and, after the temperature of the universe falls 
below the Hagedorn 
temperature following the phase transition, the fundamental 
string states become meaningless and one has to study brane 
states instead (see Refs.~\cite{stringgas1, stringgas2} for reviews). 
There 
is also an extensive body of separate literature attempting to describe 
spacetime as an entity 
emerging from still unspecified building blocks (``atoms'' or 
``molecules'' of spacetime) \cite{emergent1}-\cite{emergent9}, and another 
intriguing idea that has generated a large literature is the 
thermodynamics of spacetime \cite{spacetimethermo1, spacetimethermo2, 
spacetimethermo3}. In this view, it makes sense to speculate about 
possible 
phase transitions between different phases of spacetime, analogous to 
phase transitions in water. In the analogy, the phase change analogous to 
the liquid water-solid ice transition in freezing lakes begins when 
$a=a_0$;  
then the ice thickness grows in lakes and the universe described by 
GR expands to $a>a_0$. For lakes, it does not make sense 
to consider negative ice thickness $z<0$; in the cosmological analogy, the 
spacetime manifold view of the universe would be meaningless if $a<a_0$ 
because gravity and spacetime are in a different phase. If it carried 
through, this 
analogy would eliminate the problem of the Big Bang singularity at $a=0$ 
by stating that in the early universe (when $a(t)$ is below the critical 
value $a_0$), spacetime is in a different phase from the one we know today 
and it 
cannot be described by the classical Einstein-Friedmann equations. Pushing 
the analogy, there is no ice and $z=0$ at all times before freezing 
begins; this analogy leads to the idea of a non-classical static 
``universe'' with $a=a_0$ at all early times in a non-GR phase, with a 
static asymptotic past. This interpretation is stretched because GR would 
not be able to describe the early phase.

The analogy should be regarded with a grain of salt: for freezing lakes, 
the phase change is accompanied by a heat flux from the cooling body of 
water, which is clearly not isolated. The cosmological spacetime manifold, 
instead, is necessarily isolated, but cooling occurs because the universe 
expands.

\section{A symmetry of the equations and of their solutions}
\label{sec:3}
\setcounter{equation}{0}

A radiation fluid is a conformally invariant form of matter since the 
Maxwell equations are conformally invariant in four spacetime dimensions 
\cite{Wald, Carroll, ourreview} and it is in principle conceivable that 
some form of conformal symmetry may hold. However, this is not trivial 
because the Einstein equations are not conformally invariant even when 
conformally invariant matter sources them. Nevertheless, a symmetry 
(vaguely) related to conformal invariance does exist and it translates in 
a previously unknown symmetry of Eq.~(\ref{Vollmer}) describing the 
freezing of lakes and of its solution.

Consider the line element of a spatially flat FLRW universe and 
perform a conformal rescaling of the spacetime metric $g_{ab} \rightarrow 
\tilde{g}_{ab} =\Omega^2 \, g_{ab}$ with conformal factor $\Omega 
(x^{\alpha})$. In 
general, the new metric $\tilde{g}_{ab}$ is not a solution of the 
Einstein equations with the same form of matter (when the 
transformed equations are rewritten as effective Einstein equations, the 
effective stress-energy tensor  generated by the conformal transformation 
contains first and second covariant derivatives of $\Omega$ and does not 
satisfy any energy condition \cite{Wald, mybook}). Nevertheless, some 
residual conformal symmetry remains, as explained below.

Using conformal  time $\eta$ defined by $dt\equiv a d\eta$ \cite{Wald, 
Carroll}, the two line 
elements related by the conformal rescaling $g_{ab} \rightarrow 
\tilde{g}_{ab}=\Omega^2 \, g_{ab}$ are   
\be
ds^2 =a^2(\eta) \left( -d\eta^2 + d\vec{x}^2\right) \rightarrow 
d\tilde{s}^2  = \Omega^2 a^2 \left( -d\eta^2 + d\vec{x}^2 \right) \,.
\ee
In general, $d\tilde{s}^2$ is no longer a FLRW line element. However, a 
special 
situation arises if the conformal factor $\Omega$ depends only 
on the time coordinate $\eta$, in which case the FLRW line element remains 
FLRW with scale factor $\tilde{a}(\eta) = \Omega(\eta) a(\eta) $:
\be
d\tilde{s}^2 =\tilde{a}^2(\eta) \left( 
-d\eta^2 + d\vec{x}^2 \right) \,.
\ee
However, in general this new 
scale 
factor $\tilde{a}(\eta)$ does not satisfy 
the Einstein-Friedmann equation with a radiation fluid. An even 
more special 
situation occurs when $\Omega=a$: in this case (using $s(t)$ as a 
synonimous of $a(t)$ in the rest of this work), the transformation 
\begin{eqnarray}
s & \rightarrow & \tilde{s}= s^2 \,,\;\;\;\;\;\;\;\; s=\sqrt{\tilde{s}} 
\,,\\
&&\nonumber\\
dt &\rightarrow & d\tilde{t} = s^2 dt \,,
\end{eqnarray}
preserves the form of the radiation fluid. In fact, we have 
\be
\frac{ds}{dt}=
\frac{ds}{d\tilde{t}} \, \frac{d\tilde{t}}{dt}=\frac{\sqrt{\tilde{s}}}{2} 
\, \frac{d\tilde{s}}{d\tilde{t}}\,,
\ee
and Eq.~(\ref{Vollmerfors}) becomes
\be
\frac{d\tilde{s}}{d\tilde{t}}=\frac{ \tilde{\alpha}}{\tilde{s}} 
\ee
with $\tilde{\alpha}=2\alpha$, {\em i.e.}, it is invariant in form. The 
Friedmann equation is mapped into 
\be
\left( \frac{1}{\tilde{s}} \, \frac{d\tilde{s}}{d\tilde{t}} 
\right)^2 =\frac{ \tilde{\alpha}^2}{\tilde{s}^4} \,,
\ee
{\em i.e.}, it remains a Friedmann equation {\em for a radiation fluid}. 
The mapping 
\be
\rho=\frac{\rho_0}{s^4} \rightarrow \tilde{\rho}= \frac{\rho}{
\tilde{s}^4} = \frac{ \tilde{\rho} }{ s^8}  
\ee
leaves invariant the energy conservation equation for a radiation fluid 
\cite{mybook}
\be
\dot{\rho} +4H \rho=0 \,,
\ee
which is now mapped into
\be
\frac{d\tilde{\rho}}{d\tilde{t}} + \frac{4}{\tilde{s}} \, \frac{ 
d\tilde{s}}{d\tilde{t}} \, \tilde{\rho}=0 \,.
\ee
It is well known in the formalism of conformal transformations in FLRW 
space that, under the conformal transformation 
$\tilde{g}_{ab} 
\rightarrow \Omega^2 g_{ab}$, 
the  energy density and pressure in these spaces transforms as 
$\tilde{\rho}=\Omega^{-4} \rho$, $\tilde{P}=\Omega^{-4} P$, preserving  
the barotropic and constant equation  of state $P=w\rho$, {\em i.e.}, it 
is still $\tilde{P}=w\tilde{\rho}$ after the conformal rescaling. For a 
radiation 
fluid and $\Omega=a$, we have
\be
\rho \rightarrow \tilde{\rho}=\Omega^{-4} \, \rho = 
\Omega^{-4} \, \frac{ \rho_0}{s^4} = \frac{ \rho_0}{s^8}
= \frac{ \rho_0}{\tilde{s}^4}\,.
\ee
To summarize, the change of variables
\begin{eqnarray}
s & \rightarrow & \tilde{s}=s^2 \,,\label{symmetry1}\\
&&\nonumber\\
dt & \rightarrow & d\tilde{t}= s^2 dt \,,\label{symmetry2}\\
&&\nonumber\\
\rho & \rightarrow & \tilde{\rho}= \frac{\rho}{s^4} = \frac{\rho}{
\tilde{s}^2}  \,,\label{symmetry3}
\end{eqnarray}
leaves unchanged the form of the Einstein-Friedmann equations for a 
spatially flat, radiation-dominated, FLRW universe.

This symmetry transfers to the solution of Eq.~(\ref{Vollmer}). The 
solution of Eq.~(\ref{Vollmerfors}) 
\be
s(t)= \sqrt{ \frac{2\alpha t}{\lambda_{ice}} +\frac{1}{h^2} } 
\ee
is mapped into
\be
\tilde{s}(t)=s^2(t)=  \frac{2\alpha t}{\lambda_{ice}} +\frac{1}{h^2} \,,
\ee
but we need to express this in terms of the rescaled time using  
$\tilde{s}( \tilde{t})=\tilde{s} \left( t(\tilde{t})\right)$. By 
integrating $ d\tilde{t}/dt=\tilde{s}$, one obtains
\be
\tilde{t}=\int \tilde{s} dt = \int s^2(t)dt =\int \left( \frac{2\alpha 
t}{\lambda_{ice}} +\frac{1}{h^2} \right) dt = 
\frac{\alpha t^2}{\lambda_{ice}} + \frac{t}{h^2} \,,
\ee
where an additive integration constant has been set to zero by imposing 
that $t$ and $\tilde{t}$ have the same origin $t=\tilde{t}=0$. We can now 
solve the algebraic equation
\be
\frac{\alpha t^2}{\lambda_{ice}} + \frac{t}{h^2} -\tilde{t}=0 
\ee
for $t$, obtaining
\be
t(\tilde{t})= \frac{\lambda_{ice}}{2\alpha} \left( 
-\frac{1}{h^2} \pm \sqrt{ \frac{4\alpha \tilde{t} }{\lambda_{ice}} 
+\frac{1}{h^4} } \right)
\ee
and we choose the positive sign in front of the square root so that 
$\tilde{t} =0$ corresponds to $t=0$, consistently with what said above. 
Now
\be
\tilde{s}( \tilde{t}) =\frac{2\alpha t(\tilde{t}) }{\lambda_{ice}} 
+\frac{1}{h^2} =\sqrt{ 
\frac{2\tilde{\alpha}\tilde{t}}{\lambda_{ice}}+\frac{1}{\tilde{h}^2} } \,,
\ee
where $\tilde{\alpha}=2\alpha$ and $\tilde{h}=h^2$. In terms of the ice 
thickness,
\be
z(t) = \lambda_{ice} \left[ s(t)-\frac{1}{h} \right] =\lambda_{ice} \left[ 
\sqrt{ \frac{2\alpha t}{\lambda_{ice}} +\frac{1}{h^2}} -\frac{1}{h} 
\right]
\ee
is mapped to 
\be
\tilde{z}( \tilde{t}) = \lambda_{ice} \left[ 
\sqrt{ \frac{2\tilde{\alpha} \tilde{t}}{\lambda_{ice}} 
+\frac{1}{\tilde{h}^2}} -\frac{1}{\tilde{h}} \right] \,, 
\ee
{\em i.e.}, the solution is invariant in form under the symmetry 
operation~(\ref{symmetry1})-(\ref{symmetry3}).

\section{Conclusions}
\label{sec:4}
\setcounter{equation}{0}

The Friedmann equation (\ref{eq:11}) of FLRW cosmology, which is a first 
order constraint on the cosmic dynamics \cite{Wald} and is similar to an 
energy integral for a conservative system \cite{Goldstein}, lends itself 
to many analogies with various physical systems (usually in steady state), 
including particles in one-dimensional motion 
\cite{particle1}-\cite{particle5}, condensed matter systems 
\cite{condensed1, condensed2, condensed3, condensed4}, optical systems 
\cite{Chen15a, 
Chen15b}, glaciology \cite{Chen15a, Chen15b, facets}, equilibrium beach 
profiles \cite{beach}, and possibly other systems. Here we point out the 
analogy between the freezing of lakes and spatially flat, 
radiation-dominated FLRW universes. This analogy inspires an intriguing 
view of the very early universe as corresponding to an unknown phase of 
spacetime. If this phase is real, it cannot yet be described given our 
complete ignorance of the building blocks of spacetime and of the rules 
they 
obey, which could be determined by a complete theory of emergent gravity 
\cite{emergent1}-\cite{emergent9}, semiclassical or corpuscolar gravity 
\cite{corpuscolar1}-\cite{corpuscolar7}, quantum gravity, or quantum 
cosmology \cite{quantumgravity1, quantumgravity2, quantumgravity3, 
quantumgravity4}. In the simple analogy highlighted here, the early phase 
of the universe would remove the problem of the Big Bang singularity, 
which would simply be inappropriate to describe with the incorrect phase 
of spacetime, the same way that at high temperatures it is nonsensical 
to talk about the properties of ice that has long been converted into 
liquid water or even steam.

The analogy of the freezing of lakes with cosmology uncovers a symmetry 
property of the relevant equation for the ice thickness and of its 
solution. This symmetry is completely hidden in the treatment of freezing 
lakes and is uncovered only thanks to a residual conformal symmetry of the 
analogous radiation-dominated universe in FLRW cosmology.

Another aspect of the analogy lends itself to further development: as 
described, the above model for the freezing of lakes necessarily contains 
a simplification of the complicated processes that occur in the natural 
environment. The simplifying conditions assumed in the model, however, can 
be recreated easily in the artificial, controlled environment of a 
laboratory. Specifically, factors such as the presence/absence of 
impurities in the water and their nature and concentration, the 
atmospheric temperature and its variation in time, the temperature at the 
bottom of the ``lake'' (in practice, a deep tank), the lack of winds, and 
the depth of the ``lake'' can all be controlled in a laboratory setting. 
The equipment necessary to conduct an analogue gravity experiment bsed on 
the physics of water is common in cold laboratories studying snow and ice, 
while the equipment required is not sophisticated in comparison with that 
used in conventional analogue gravity in which black holes, cosmological 
spacetimes, and curved space phenomena such as Hawking radiation, 
superradiance, and false vacuum decay require the use of Bose-Einstein 
condensates \cite{condensed1, condensed2, condensed3, condensed4, 
falsevacuum}, ultracold atoms \cite{ultracold}, optical systems 
\cite{optical1, optical2}, or at least very finely controlled water flows 
and vortices ({\em e.g.}, \cite{flow1, flow2, flow3, flow4}). Likewise, 
the experimental study of the analogy between freezing lakes and cosmology 
would require a much simpler laboratory setting than it would be necessary 
to study the analogy between cosmology and large geological systems such 
as glaciers and beaches, which also exhibit analogies with cosmology 
\cite{facets, beach}.

\begin{acknowledgments}

This work is supported, in part, by Bishop's University and by the Natural 
Sciences \& Engineering Research Council of Canada (Grant No. 2016-03803).

\end{acknowledgments}

\end{document}